\documentclass[twocolumn,showpacs,preprintnumbers,amssymb]{revtex4}
\usepackage{graphicx}
\usepackage{dcolumn}
\usepackage{amsmath}
\usepackage{amssymb}
\begin{document}
\preprint{APS/V2-05/10/04}

\title{Kolmogorov and Irosnikov-Kraichnan scaling in the anisotropic turbulent solar wind.}
\author{Sandra C. Chapman}\email{S.C.Chapman@warwick.ac.uk}
\author{Bogdan Hnat}

 \affiliation{Centre for Fusion, Space and Astrophysics, Physics Department, University of Warwick, Coventry, CV4 7AL, UK.}

\date{\today}

\begin{abstract}
Solar wind turbulence is dominated by Alfv\'{e}nic fluctuations
but the power spectral exponents somewhat surprisingly evolve
toward the Kolmogorov value of $-5/3$, that of hydrodynamic
turbulence. We show that at 1AU the turbulence decomposes linearly
into two coexistent components perpendicular and parallel to the
local average magnetic field. The first of these is consistent
with propagating Alfv\'{e}n wavepackets and shows the scaling
expected of Alfv\'{e}nic turbulence, namely Irosnikov- Kraichnan.
The second shows Kolmogorov scaling which we also find in the
number and magnetic energy density, and Poynting flux.

\end{abstract}
  \maketitle

The solar wind provides a unique laboratory for the study of
Magnetohydrodynamic (MHD) turbulence with a magnetic Reynolds
number estimated to exceed $10^5$ in the solar
wind\cite{matthprl}. In-situ satellite observations of bulk plasma
parameters strongly suggest the presence of turbulence via the
statistical properties of their fluctuations\cite{cytu,unsolved}.
Quantifying these fluctuations is also central to understanding
both the transport of solar energetic particles and galactic
cosmic rays within the heliosphere, and solar wind evolution with
implications for the mechanisms that accelerate the wind at the
corona.

The observed fluctuations in the solar wind present a complex
mixture of hydrodynamic and Alfv\'{e}nic signatures. Alfv\'{e}nic
fluctuations dominate the power in these fluctuations and are
observed propagating away from the sun implying solar origin (e.g.
\cite{horbury}). However the power spectra
\cite{cytu,unsolved,horbury,bale} suggest an exponent evolving
toward the Kolmogorov\cite{k41} (hereafter K-41) value of $\sim
5/3$, that of hydrodynamic
 turbulence. This is paradoxical since for ideal
 MHD the turbulent cascade is expected to be mediated via Alfv\'{e}n
wavepackets suggesting an exponent of $\sim -2/3$, that of
Irosnikov and  Kraichnan\cite{IK} (hereafter IK). Intervals can be
found where different magnetic field and velocity components
simultaneously exhibit K-41 and IK scaling\cite{veltri,bershpre},
indeed, these phenomenologies can be difficult to distinguish in
low order moments\cite{carbprl}. The flow is also observed to be
intermittent, this has been suggested to account for the
'anomalous' $-5/3$ scaling in the power spectra in terms of
incompressible MHD, rather than hydrodynamic,
phenomenology\cite{carbcasprl}. Alfv\'{e}nic fluctuations, when
isolated by the use of Elsasser variables (see
e.g.\cite{horbury}), and decomposed by considering different
average magnetic field orientations that occur at different times,
are found to be multicomponent\cite{matg1990}, and
coupled\cite{milprl}. This picture, of an essentially
incompressible, multicomponent Alfv\'{e}nic
turbulence\cite{matg1990,horbury} suggests that a significant
population of Alfv\'{e}nic fluctuations evolve to have wavevectors
almost perpendicular to the background magnetic field, leading to
a 'fluid- like' (in the sense of K-41) phenomenology, and the
$-5/3$ power spectral slope. However, fluctuations in solar wind
density are not simply proportional to that in magnetic
field\cite{spanglerpop} and show nontrivial
scaling\cite{cytu,hnatpre} that suggests that the turbulence is
compressible\cite{hnatprl}. The role of compressibility is thus an
open question. An important corollary is that the full behaviour
cannot be captured by models which describe the observed
Alfv\'{e}nic properties in terms of fluctuating coronal fields
that have advected passively in the expanding solar
wind\cite{giac}.

Here, we will quantify the interplay between K-41 and IK
phenomenologies in the turbulent solar wind. We can discuss the
statistical properties of fluctuations in some variable of the
flow, such as velocity, magnetic field, or density, by considering
ensemble averages. Fluctuations in the velocity field can be
characterized by the difference in some component, or in the
magnitude, $\delta v=v(r+L)-v(r)$ at two points separated by
distance $L$. The dependence of $\delta v$ upon $L$ is determined
in a statistical sense through the moments $<\delta v^p>$, where
$<...>$ denotes an ensemble average over $r$. Statistical theories
of turbulence then anticipate scaling $<\delta v^p_L>\sim
L^{\zeta(p)}$.

 Kolmogorov's 1941 theory for
hydrodynamic flows\cite{k41} essentially follows from dimensional
analysis. A fluctuation $\delta v$ arising from a transient
structure in the flow with characteristic lengthscale $L$, and
timescale $T$, transfers kinetic energy $\delta v^2$ implying an
energy transfer rate $\epsilon_L\sim \delta v^2/T \sim \delta
v^3/L$. If the statistics of the fluctuations in the energy
transfer rate are independent of $L$, its $p$ moments
$<\epsilon_L^p>\sim \epsilon_0^p$ where the constant $\epsilon_0$
is the average rate of energy transfer. This gives the K-41
scaling $<\delta v^p_L>\sim L^{p/3}$. In practice, hydrodynamic
flows are found to deviate from this simple scaling. This
intermittency\cite{frisch} is introduced through a lengthscale
dependence of the fluctuations in energy transfer rate so that
$<\epsilon^p_L>\sim \epsilon_0^p(L/L_0)^{\mu(p)}$, where $L_0$ is
some characteristic lengthscale and $\mu(p)$ is the intermittency
correction. The scaling for the moments then becomes $<\delta
v_L^p>\sim L^{\zeta(p)}$ with the K-41 exponents
$\zeta(p)=p/3-\mu(p/3)$. For incompressible MHD turbulence,
Alfv\'{e}nic phenomenology mediates the cascade, introducing an
additional characteristic speed, the Alfv\'{e}n speed $v_A$. The
above dimensional argument then gives an energy transfer rate
$\epsilon_L\sim (\delta v^2/T)(\delta v/v_A)\sim \delta v^4/L$,
which is just that proposed by Iroshnikov and Kraichnan\cite{IK},
so that $<\delta v^p> \sim L^{\zeta(p)}$ now with
$\zeta(p)=p/4-\mu(p/4)$.

The experimental study of turbulence then centres around
measurement of the scaling exponents, the $\zeta(p)$. A full
description requires the (difficult to determine) intermittency
correction, the $\mu(p)$. However, if the system is in a
homogeneous steady state, the average energy transfer rate is
uniform so that $<\epsilon_L>=\epsilon_0$ and $\mu(1)=0$ giving,
for K-41 $\zeta(3)=1$, and for IK, $\zeta(4)=1$, independent of
the intermittency of the flow. A determination of the lower order
moments that is sufficiently accurate to distinguish these two
cases is possible for in- situ observations of the solar wind and
we present this here.
\begin{figure}
\resizebox{\hsize}{!}{\includegraphics{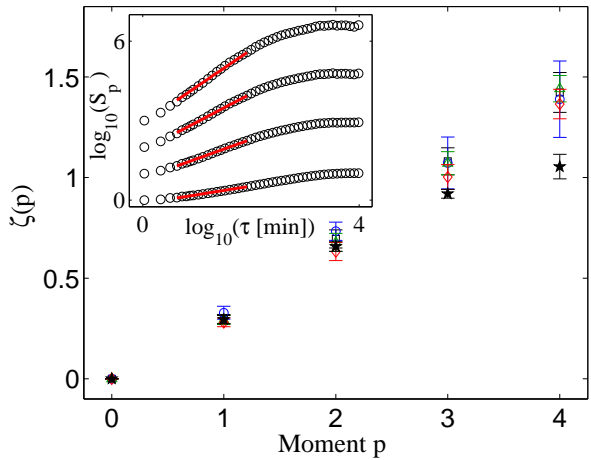}}
\caption{\label{fig1}Structure function analysis of solar wind
density fluctuations. Inset: structure functions versus
differencing interval $\tau$ (the traces are offset for clarity).
Main plot: scaling exponents computed from the raw data ($\star$),
and applying an upper limit to fluctuation size of $20
\sigma(\tau) (\Box)$, $15 \sigma(\tau)(\circ)$, $10 \sigma(\tau)
(\bigtriangleup)$ and $5 \sigma(\tau) (\diamond)$.}
\end{figure}
These observations are typically time series from a single
spacecraft so that the ensemble averages that we will consider
will be over time rather than over space, the spatial separation
$L$ above being replaced by a time interval $\tau$- the Taylor
hypothesis\cite{matthprl}. Consistent with almost all experimental
studies of turbulence we consider generalized structure functions
of a given parameter $x$: $S_p(\tau)=\left<\mid
x(t+\tau)-x(t)\mid^p\right>$. Solar wind monitors such as the ACE
spacecraft spend several-year long periods in orbit about the
Lagrange point sunward of the earth.
 We analyse $64 s$ averaged plasma parameters from ACE for
 the interval 01/01/1998 - 12/31/2001, this consists
 of $\sim 1.6 \times 10^6$ samples and is dominated by slow solar wind. Figure 1 shows the procedure
 for extracting the scaling exponents from the data. The inset panel shows the structure functions
  of fluctuations in the density versus differencing interval
     for $p=1-4$. There is a
      scaling range for timescales of minutes up to a
few
     hours, the timescale for large scale coherent structures. This scaling range
      has been shown to extend up to almost three orders of magnitude via Extended
       Self Similarity (ESS)\cite{hnatprl}.  The scaling exponents, that is, the $\zeta(p)$,
        where $S_p(\tau)\sim \tau^{\zeta(p)}$, are the
       gradients of these scaling regions, and these are shown in the main plot.
        The error bars provide an estimate of the uncertainty in the gradients
         of the fitted lines (linear regression error). Finite, experimental
         data sets include a small number of extreme events which have poor
          representation statistically and may obscure the scaling properties
          of the time series. One method\cite{veltri,npg}  for excluding these rare events
           is to fix a (large) upper limit on the magnitude of fluctuations
           used in computing the structure functions. Importantly, this limit
            is varied with the temporal scale $\tau$ to account for the growth of
            range with $\tau$ in the time series. The figure shows the exponents
            computed for a range of values for this upper limit
            $[5,20]\sigma(\tau)$, where
            $\sigma(\tau)=S_2^{1/2}$.
We see that the scaling exponents are insensitive to the value of
the upper limit once a limit is applied and the rare large events
are removed. Above $10\sigma(\tau)$ this process eliminates less
than 1\% of the data points.

\begin{figure}
\resizebox{\hsize}{!}{\includegraphics{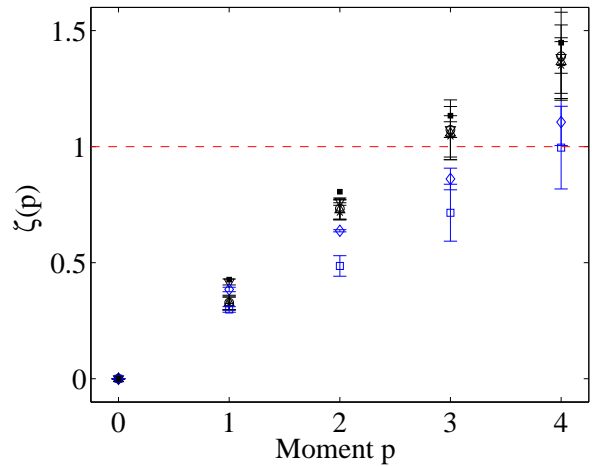}}
\caption{\label{fig2} Scaling exponents $\zeta(p)$ versus $p$ for
solar wind quantities, in blue, magnitudes of velocity $v (\Box)$
and magnetic field $B(\diamond)$, in black, number density
$\rho(\circ)$, number flux $\rho v (\bigtriangleup)$ and momentum
flux $\rho v^2 (\diamond)$, magnetic energy density
$B^2(\bigtriangledown)$ and Poynting flux $vB^2 (\blacksquare)$.
Note that $\zeta(4)\approx 1$  and $\zeta(3)\approx 1$
respectively for these groups of quantities.}
\end{figure}

We now compare the scaling exponents for different scalar
quantities in the solar wind flow, for structure functions up to
$p=4$. In Figure 2 we show the scaling exponents for fluctuations
in the magnitude of velocity $v$ and magnetic field $B$ (blue
symbols), along with those for the number density $\rho$, the
magnetic energy density $B^2$, the flux density $\rho v^2$ and the
Poynting flux in the MHD limit $v B^2$ (black symbols). The
scaling exponents for these low order structure functions are
determined with sufficient precision that we can see that in the
case of the velocity and magnetic field magnitudes, the $\zeta(4)$
exponents are clustered about unity, whereas for the other
quantities, the $\zeta(3)$ exponents are clustered about unity.
Thus the fluctuations in velocity and magnetic field magnitude
appear to be dominated by Alfv\'{e}nic processes, in the sense of
IK phenomenology, whereas the fluctuations in the magnetic field
energy, plasma and flux densities are hydrodynamic- like, in the
sense of K-41. Intriguingly, these two phenomenologies are
coexistent.
\begin{figure}
\resizebox{\hsize}{!}{\includegraphics{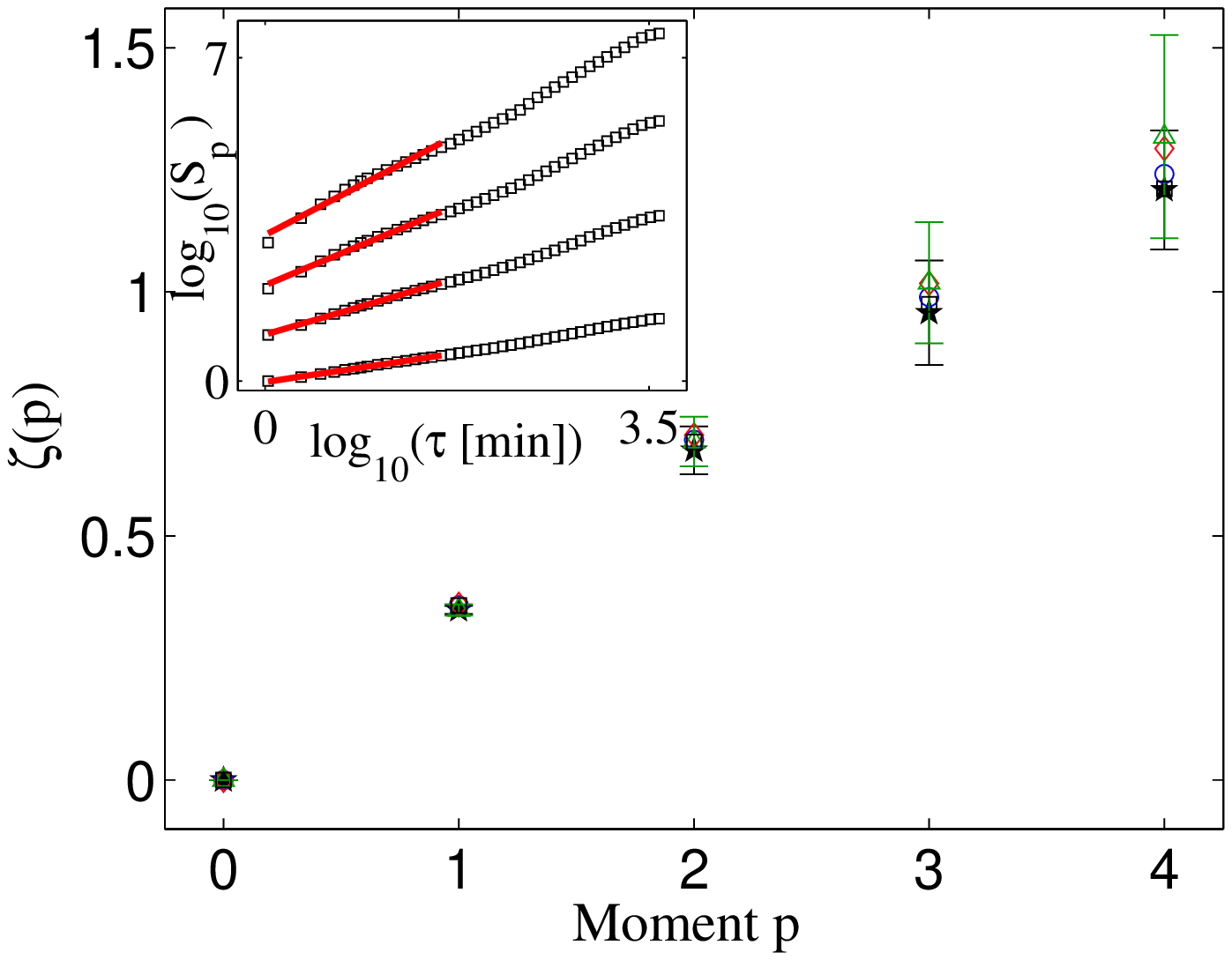}}
\caption{\label{fig3} Structure function analysis of $<\mid \delta
{\bf{v}}\cdot\hat{\bf{b}}\mid^p>$ . Inset: structure functions
versus differencing interval (the traces are offset for clarity).
Main plot: scaling exponents computed from the raw data ($\star$),
and applying an upper limit to fluctuation size of $20
\sigma(\tau) (\Box)$, $15 \sigma(\tau)(\circ)$, $10 \sigma(\tau)
(\bigtriangleup)$ and $5 \sigma(\tau) (\diamond)$.}
\end{figure}
\begin{figure}
\resizebox{\hsize}{!}{\includegraphics{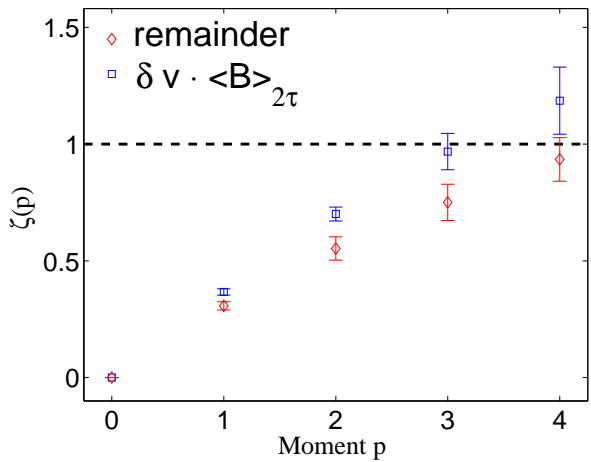}}
\caption{\label{fig4} Scaling exponents $\zeta(p)$ versus $p$ for
the structure functions of $<\mid \delta {\bf{v}} \cdot
\hat{\bf{b}} \mid^p> (\Box)$ and of the remaining signal
($\diamond$). Note that $\zeta(3) \approx 1$ and $\zeta(4) \approx
1$ respectively for these quantities.}
\end{figure}

We now introduce an operation that 'filters out' one of the
relevant physical processes, namely,  Alfv\'{e}nic fluctuations.
We exploit the property that the full non- linear MHD equations
support large scale Alfv\'{e}nic fluctuations which share a basic
property of Alfv\'{e}n waves- that the velocity perturbation is
perpendicular to the background magnetic field. In the turbulent
flow, the magnetic field also fluctuates, but we can consider a
local background value by constructing a running average of the
vector magnetic field over the timescale $\tau'$. For each
interval over which we obtain a \emph{difference} in velocity
$\delta \bf{v}=\bf{v}(t+\tau)-\bf{v}(t)$ we also obtain a vector
\emph{average} for the magnetic field direction
$\hat{\bf{b}}=\bar{\bf{B}}/\mid \bar{\bf{B}}\mid$ from a vector
sum of all the observed vector values between $t$ and $t+\tau'$,
$\bar{\bf{B}}(t,\tau')=\bf{B}(t)+... +\bf{B}(t+\tau')$, with
$\tau'$ centred on $\tau$. We choose the interval $\tau'=2 \tau$
here as the minimum (Nyquist) necessary to capture wavelike
fluctuations. Velocity differences $\delta \bf{v}$ which are
Alfv\'{e}nic in character will then have the property that the
scalar product $\delta \bf {v} \cdot \hat{\bf{b}}$ will vanish.
This condition filters out \emph{all} those fluctuations which
generate a velocity displacement perpendicular to the local
magnetic field, and is thus less restrictive than the
Elsasser\cite{horbury} variables which select propagating pure
Alfv\'{e}n waves. In Figure 3 we plot the structure functions of
the quantity $\delta v_\parallel=\delta \bf {v} \cdot
\hat{\bf{b}}$, that is,
$S_p=<\mid\delta{\bf{v}}\cdot\hat{\bf{b}}\mid^p>$ versus $\tau$,
(inset) and the corresponding scaling exponents, the $\zeta(p)$
for the region where $S_p \sim \tau^{\zeta(p)}$ (main plot)
generated in the same way as in Figure 1. In Figure 4 we compare
these exponents with those obtained for the remaining signal, that
is, $\delta v_\perp=\sqrt (\delta\bf{v}\cdot\delta\bf{v}-(\delta
\bf {v} \cdot \hat{\bf{b}})^2)$. Remarkably, both these quantities
show a clear scaling range (which we will verify) with scaling
exponents $\zeta(3)$ and $\zeta(4)$ close to unity for $\delta
v_\parallel$ and $\delta v_\perp$ respectively.

\begin{figure}
\resizebox{\hsize}{!}{\includegraphics{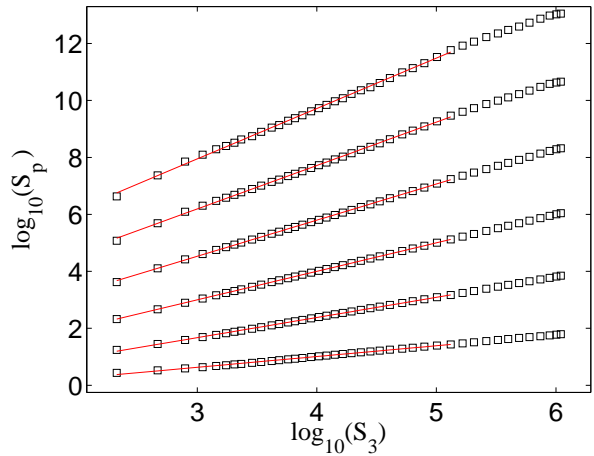}}
\caption{\label{fig5} Structure functions $S_p$ versus $S_3$ for
$p=1-6$ for $S_p=<\mid\delta{\bf{v}}\cdot\hat{\bf{b}}\mid^p>$. The
traces are offset for clarity.  }
\end{figure}
This result is consistent with the fluctuations in velocity being
a simple linear superposition  that are close to (i) parallel to
the local background magnetic field and sharing hydrodynamic- like
scaling, that is K-41, with that of the number, magnetic energy
and flux densities, and (ii) perpendicular to the local background
magnetic field with the scaling expected of Alfv\'{e}nic
turbulence in the sense of IK. These results provide the first
unambiguous ordering of the data with respect to hydrodynamic
(i.e.Kolmogorov-like), and Alfv\'{e}nic (i.e. Irosnikov-
Kraichnan-like) phenomenology in different parameters observed at
a single location in the solar wind.

\begin{figure}
\resizebox{\hsize}{!}{\includegraphics{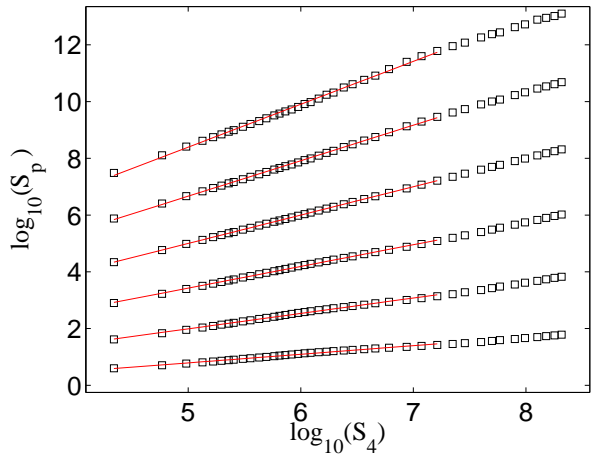}}
\caption{\label{fig6} Structure functions $S_p$ versus $S_4$ for
$p=1-6$ for $S_p=<\mid
\sqrt(\delta{\bf{v}}\cdot\delta{\bf{v}}-(\delta {\bf{v}} \cdot
\hat{\bf{b}})^2) \mid^p>$. The traces are offset for clarity. }
\end{figure}

We verify that these quantities indeed show an extended scaling
region by means of ESS\cite{benzi}. If the scaling is such that
the   $S_p \sim S_q^{\zeta(p)/\zeta(q)}$ then a plot of $S_p$
versus $S_q$ will reveal the range of the underlying power law
dependence with $\tau$. If, as here, one of the $\zeta(p)$ are
close to unity, the ESS plot will in addition provide a better
estimate of the $\zeta(p)$. Figures 5 and 6 show $S_p$ versus
$S_3$ for $\delta v_\parallel$ and versus $S_4$ for $\delta
v_\perp$ respectively, and we see that there is scaling over
several orders of magnitude. The slopes of these plots imply
exponents that are multifractal, that is, quadratic in $p$, and
distinct for the two cases.

 The
characteristic nature of solar wind turbulence is revealed to be a
coexistence of two signatures. The first of these is consistent
with Alfv\'{e}nic turbulence in the sense of IK mediated by
Alfv\'{e}n wavepackets propagating parallel to the magnetic field.
The second, which shows K-41 scaling, has a compressive component
and could couple nonlinearly into strongly oblique, almost non
propagating Alfv\'{e}nic fluctuations. This clearly elucidates the
previously proposed multicomponent nature of solar wind turbulence
and suggests one of two scenarios. The first of these is that the
turbulent solar wind is comprised of two weakly interacting
components- one from the process that generates the solar wind at
the corona and the other that evolves in the high Reynolds number
flow. Given the evidence for outward propagating Alfv\'{e}n waves
and evolution toward K-41 scaling, these correspond to IK, and
K-41 phenomenology respectively and our result yields an important
insight into the physics of solar wind generation. Alternatively,
the coexistence of the two components is characteristic of the
anisotropic nature of compressible MHD turbulence in the presence
of a background field, in which case this determination of their
scaling properties points to an important modification of theories
of MHD turbulence.

\section{Acknowledgment}
The authors thank G. Rowlands for discussions, the ACE Science
Centre for data provision and the PPARC for support.

\end{document}